\documentclass[aps,amsmath,showpacs,twocolumn]{revtex4-1}

\usepackage{graphics,epsfig}

\begin{document}


\title{Model-independent form factor relations  at large $N_c$ }

\author{Thomas D. Cohen}
\email{cohen@physics.umd.edu}

\author{Vojt\v{e}ch Krej\v{c}i\v{r}\'{i}k}
\email{vkrejcir@umd.edu}

\affiliation{Maryland Center for Fundamental Physics, Department of Physics, \\
 University of Maryland, College Park, MD 20742-4111}

\begin{abstract}
In this paper  a model-independent relation which holds for the long distance part of the  Fourier transform of the electromagnetic form factors of the nucleon in the large $N_c$ and chiral limits is demonstrated.  This relation was previously conjectured based on the fact that it emerged in {\it all} semiclassical chiral models independent of the details of the model.  Here we show that the result is, in fact, model independent by deriving it directly in large $N_c$ chiral perturbation theory which is known to capture the long distance behavior of the form factors.   The relation is valid when the large $N_c$ limit is formally taken before the chiral limit.  A new relation is derived for the case where the chiral limit is taken prior to the large $N_c$ limit.

\end{abstract}


\maketitle

\section{Introduction}

Low-energy phenomena related to strong interaction are ultimately traceable directly from Quantum Chromodynamics (QCD), the fundamental theory of strong interactions. It is often difficult to do this since the theory is strongly coupled and the conventional perturbative expansion in powers of a coupling constant, i.e., around a non-interacting theory, is not applicable.  In order to gain insight into the phenomena it is often useful to use  alternative expansion schemes such as the chiral expansion about the massless quark limit, and the $1/N_c$ expansion about the the large $N_c$ limit.

The number of colors  $N_c$ is a hidden parameter of QCD, which is  built upon the gauge $SU(N_c)$ symmetry with $N_c=3$. It was shown \cite{tHooftNC, WittenNC} that many aspects of QCD simplify substantially in the limiting case of infinite number of colors $N_c \rightarrow \infty$. Thus it makes sense to study the theory in this limit and to establish an expansion in the powers of $1/N_c$ around this 'simpler' theory.
The chiral expansion is based on the fact that the mass of the pion is much lower than all other scales \cite{Weinberg}. Moreover, in the chiral limit, $m_{\pi}\rightarrow 0$, QCD possesses a new symmetry, the chiral symmetry, which, again, simplifies the problem substantially.
Consequently, it is a promising approach to develop models of QCD in these two limits.
Even though these limits do not completely describe the real world, they are believed to capture many of its main (at least qualitative) features. Moreover, one can develop a systematic procedure how to include the corrections in the powers of either $1/N_c$ or  $m_{\pi}$.  It is important to note that the double limit is not uniform for certain observables and that the leading behavior may depend on the ordering of the two limits \cite{CohenBroniowski,Cohen95,Cohen96}.

 It was pointed out by Witten \cite{WittenNC} that in the large $N_c$ limit, QCD becomes a weakly interacting theory of mesons with baryons emerging as soliton-like configurations of meson fields.
A class of baryon models based on this observation plus approximate chiral symmetry has been developed, namely chiral soliton models treated semi-classically.  The large $N_c$ limit of QCD is encoded in the very core of these models and its semiclassical treatment; their degrees of freedom are weakly interacting mesons forming soliton-like solutions identified with baryons. The issue of chiral limit, and following chiral symmetry, is somewhat more subtle; it is imposed at a later stage as a constraint on the dynamics of  meson fields. Such observations can be formulated more rigorously: chiral soliton models when treated semiclassically are models based on large $N_c$ and chiral limits of QCD with $N_c \rightarrow \infty$ taken first.  An example of such  a model is the famous Skyrme model \cite{Skyrme1, AdkinsNappiWitten}. Here, baryons are identified with the quantum states of collective motions of Skyrmions, the hedgehog configurations of meson fields.

In principle, one could develop  an infinite number of models based on large $N_c$ and chiral limits of QCD.  Of course, there is more to modeling QCD than simply getting the large $N_c$ and chiral behavior correct and thus the class of models can differ substantially in dynamical detail.  Still, it is important to develop a method how to check whether the large $N_c$ and chiral physics are encoded correctly in these models.   The obvious tools are various model-independent relations. There is a large class of model-independent constraints based on consistency relations for large $N_c$ limit \cite{GervaisSakita84-1, GervaisSakita84-2, DashenManohar93-1, DashenManohar93-2, DashenJenkinsManohar94, DashenJenkinsManohar95}.  Such consistency relations can greatly constrain the chiral behavior associated with the longest distance behavior of the system \cite{CohenBroniowski,Cohen95,Cohen96} provided that it is understood that these relations hold when the large $N_c$ limit is taken prior to the chiral limit.  Any viable model based on the large $N_c$ and chiral limits needs to reproduce these relations.  It is worth noting that semiclassical chiral soliton models reproduce all known model-independent relations of this type, independently of all the detailed dynamics of the model.  Indeed, many of these relations were first discovered in the context of the Skyrme model when it was noted by Adkins and Nappi \cite{AdkinsNappi} that certain relations held regardless of the parameters of the model.

Recently, there has been a wide interest in using gauge/gravity duality in the construction of holographic baryon models \cite{PomarolWulzer1, PomarolWulzer2, PanicoWulzer, SakaiSugimoto1, SakaiSugimoto2, 21, HSakaiSugimotoY, HSakaiSugimoto, HasSakaiSugimoto2}. These models also  implicitly contain large $N_c$ limit in the very core of their construction.
The details of these models look quite different from the traditional 4D soliton-based models---if for no other reason the models are in five dimensions.
The boundary values of some of the bulk fields act as
sources for operators with meson quantum numbers in the field theory,
and baryons are modeled as quantum states of topologically nontrivial configurations of these bulk fields.  Because these models are so different from conventional soliton models and are rather nontrivial to implement, it is particularly important to test whether they have, in fact, correctly encoded the chiral and large $N_c$ scaling behavior.   To do this a new class of model-independent relation needs to be formulated.  The reason for this is that  the chiral behavior associated with long distance behavior in the typical model-independent relation simply fixed the coefficient of how certain quantities such as charge radii diverge as $m_\pi \rightarrow 0$.  However, to date many of the treatments of these models have only been done for $m_\pi=0$ and hence one cannot use the behavior of how these quantities diverge with $m_\pi$.  Instead of using the $m_\pi$ dependence to probe the long distance behavior, one can use the long distance behavior of the Fourier transform of the electromagnetic form factors as a model-independent probe \cite{ChermanCohenNielsen}.

This new type of model-independent relation has already proven to be of great value in studying holographic soliton models.  While  a ``bottom up'' phenomenological model \cite{PomarolWulzer1, PomarolWulzer2, PanicoWulzer} could be shown to satisfy the new  model-independent relation, indicating that it had correctly captured the large $N_c$ and chiral behavior, the treatment of solitons as instantons \cite{SakaiSugimoto1, SakaiSugimoto2, 21, HSakaiSugimotoY, HSakaiSugimoto, HasSakaiSugimoto2} in the ``top down'' model of Sakai and Sugimoto \cite{SakaiSugimoto1} failed to satisfy the relation indicating that something was seriously wrong with the approach despite the claimed phenomenological successes of the approach. It was recently  shown by Cherman and Ishii \cite{ChermanIshii} that the underlying reason for this appears to be due to a failure of the
flat-space
instanton approximation. This may have important phenomenological consequences since it was this approximation which led to a vector-meson dominance picture \cite{SakaiSugimoto2} and the one-meson exchange picture on nucleon-nucleon forces \cite{HasSakaiSugimoto2}.

A particularly useful relation of this type is for a ratio of the product of the isovector and isoscalar position-space electric  form factors to the product of the magnetic nucleon form factors in the long distance limit:
\begin{equation}
\lim\limits_{r \rightarrow \infty} \, \frac{  r^2 \,\widetilde{G}_E^{I=0} \,\widetilde{G}_E^{I=1}}{\widetilde{G}_M^{I=0} \,\widetilde{G}_M^{I=1}}  \,=\, {18},
\label{magicalratio}
\end{equation}
where the position-space isoscalar and isovector electric and magnetic form  factors $\widetilde{G}_{E,M}$ are Fourier transformed momentum-space  electric and magnetic form factors $G_{E,M}$  \cite{ErnstSachs}.

The remarkable thing about   this particular ratio (\ref{magicalratio}) is that \emph{all} low-energy constants, normalization of currents, and various sign or Fourier transform conventions cancel. Thus, the final result depends only on a given power of radius $r$ (easily deduced from dimensional analysis) multiplied by a constant. This constant, 18, is a universal model-independent quantity that must be satisfied in all large $N_c$ chiral models.

 We note in passing that the number 18 is particularly significant.  The Hebrew characters for 18 corresponding to the word ``chai'' which means life and thus 18 is considered to be a lucky number in the Jewish tradition.   Moreover there are 18 chapters in the Bhagavad Gita.  Finally we note that 18 is the drinking age in most civilized countries including the Czech Republic.

The relation in Eq. (\ref{magicalratio})  {\it does} depend on the order of limits and is valid if the large $N_c$ limit is taken prior to the chiral and large $r$ limits.   Equation (\ref{magicalratio}) was originally \cite{ChermanCohenNielsen} derived in a similar spirit to Adkins and Nappi \cite{AdkinsNappi}, namely using the Skyrme model to find a truly model-independent relation.  It turns out that Eq. (\ref{magicalratio}) is true in {\it any} Skyrme model, independent of the details of the Lagrangian or the number of degrees of freedom in the problem provided that the model has a well-defined chiral limit in which the pions emerge as Goldstone bosons and provided that the model is treated semi-classically.   This seems to be compelling evidence that the result is truly model independent since in all known cases where a chiral result is derived from the Skyrme model, and turns out to be completely independent of model details, have also turned out to be the result of large $N_c$ chiral perturbation theory which is known to correctly describe the longest distance quantities in QCD.   However, it is important to verify Eq.  (\ref{magicalratio}) in a truly model-independent way directly from large $N_c$ chiral perturbation theory.  Doing so is the principal purpose of this paper.  We will also explore what happens to the longest distance behavior of the form factors when the ordering of limits is changed so that the chiral limit is taken prior to the large $N_c$

 The long distance hadronic physics in QCD is dominated by the pion cloud, since a pion is the lightest particle available. Moreover, in the baryon sector, unlike in the meson one, the pion loops contribute at leading order in $1/N_c$ expansion \cite{CohenLeinweber}, so the dominant diagrams to consider to describe the longest distance behavior will consist of currents connected to the pion in loops containing the fewest possible  number of pions.
Additional simplification comes from the fact that baryon mass is of order $N_c$ and thus parametrically large compared to the pion mass. It allows us to neglect the recoil of the baryons and treat them non-relativistically. In other words, we will work in the heavy baryon $\chi PT$.

Note that we work in the combined large $N_c$ and chiral limit. In this case, a subtlety arises for observables sensitive to the dynamics of the pion. These two limits generally do not commute and therefore the ordering of limits matter \cite{CohenBroniowski, Cohen96}.
In our paper we  start by considering  the following ordering:  $N_c\rightarrow\infty$ limit first, $m_{\pi}\rightarrow 0$ limit second. It is exactly the ordering that both soliton models of baryons  (both four dimensional  and holographic   models)  use implicitly.

The role of the large $N_c$ limit in $\chi PT$ is two-fold. First, it eliminates diagrams that are  suppressed by factors $1/N_c$. Second, it forces one to include $\Delta$-isobar in the calculation. The large $N_c$ consistency relations require nucleon and $\Delta$ to be degenerate (in general the whole tower of $I=J$ isobars) at large $N_c$.  More specifically, the mass difference $\Delta=m_{\Delta}-m_N$ is of order $1/N_c$ and serves as a new low-energy constant.  Thus, at lowest order in large $N_c$, $\chi$PT depends on the constants  $g_A$, $f_{\pi}$,  $m_{\pi}$ and $\Delta$.   The need to include the $\Delta$ increases the number of Feynman diagrams contributing and has far-reaching consequences.

It was shown \cite{CohenBroniowski, Cohen96} that for isoscalar-scalar ($\widetilde{G}_E^{I=0}$) and isovector-vector ($\widetilde{G}_M^{I=1}$) channels, $\Delta$ in the intermediate state only leads to a multiplicative factor. However, in the isoscalar-vector ($\widetilde{G}_M^{I=0}$) and isovector-scalar ($\widetilde{G}_E^{I=1}$) channels,  the amplitudes corresponding to individual diagrams subtract exactly in the leading order in $1/N_c$ (where $\Delta =0$), and one must take into account the $N$-$\Delta$ mass splitting. Consequently $\widetilde{G}_M^{I=0}$ and $\widetilde{G}_E^{I=1}$ are proportional to $\Delta$ (and are of order $1/N_c$). Note that in the relation (\ref{magicalratio}) one of these quantities is in the numerator and one of them in the denominator, so that the dependence on $\Delta$ cancels.

\section{Position-space form factors}

The calculation of form factors is performed  in the language of quantum field theory. Thus,
  we need the Feynman rules for the following vertices: photon and two pions, photon and three pions, pion with baryons in all four possible combinations of incoming and outgoing nucleon and $\Delta$, as well as propagators of intermediate pions, nucleons, and $\Delta$s.
Details of the calculation are in the appendix, so only basic assumptions, building blocks and results are presented here.

The Feynman rules can be obtained from the standard $\chi$PT Lagrangian \cite{BernardKaiserMeissner}.
The interaction of photon with two pions is crucial for the isovector coupling of photon to nucleon (see the $\epsilon_{a3b}$). The Feynman rule for the vertex reads:
\begin{equation}
 \epsilon_{a3b} \, A_{\mu} \, (p_a^{\mu} + p_b^{\mu}),
\label{vertexFPPtext}
\end{equation}

The coupling of photon with three pions relates to QCD anomaly \cite{Witten82}. It is crucial for the isoscalar piece of the baryon current (see the $\epsilon_{abc}$).
The Feynman rule for the corresponding vertex is:
\begin{equation}
\frac{1}{12 \pi^2 f_{\pi}^3} \, \epsilon_{abc} \, \epsilon^{\mu \nu \kappa \lambda }   A_{\mu} \, p_{a{\nu}} \, p_{b{\kappa}} \, p_{c{\lambda}}.
\label{vertexFPPPtext}
\end{equation}

The interaction of a pion with a baryon  is of the vector-isovector form and is derivatively coupled. The Feynman rule for incoming baryon $B$ and outgoing baryon $B'$ reads:
\begin{equation}
\frac{g_A}{2 f_{\pi}  } \,\,\sqrt{\frac{2J^{(B')}+1}{2J^{(B)}+1}} \,\, \tau^{(BB')}_a \,\, \sigma^{(BB')}_i \, p_i,
\label{vertexPBBtext}
\end{equation}
In (\ref{vertexPBBtext}), $\tau^{BB'}$ and $\sigma^{BB'}$ are operators acting in the isospin and in the spin space, respectively. They are a generalization of Pauli matrices, which appear in the pion-nucleon-nucleon vertex in standard $\chi$PT \cite{BernardKaiserMeissner}. The form of the couplings is determined by the consistency relations of large $N_c$ QCD \cite{DashenManohar93-1, DashenManohar93-2}. Simply, the interaction is of the same form as standard pion-nucleon-nucleon (vector-isovector) where ordinary $2\times 2$ Pauli matrices ($\tau$ and $\sigma$) are replaced by the properly normalized $(2J'+1)\times (2J+1)$ matrices ($\tau^{BB'}$ and $\sigma^{BB'}$) whose elements are given by the Clebsch-Gordan coefficients (see appendix for more details).

The other two low-energy constants ($m_{\pi}$ and $\Delta$) enter the calculation in propagators of intermediate state particles. Since we intend to set $m_{\pi} \rightarrow 0$ in the end (chiral limit) we must use the full relativistic pion propagator ($k$ being propagating 4-momentum):
\begin{equation}
\Delta^{\pi}(k)= \frac{i}{k^2-m_{\pi}^2+ i \epsilon} .
\label{propPtext}
\end{equation}
The heavy baryon approximation, which is justified by the fact that we work in the large $N_c$ limit, allows us to treat baryons non-relativistically:
\begin{equation}
\Delta^{N}(k)= \frac{i}{k^0+i \epsilon}\,,\,\,\Delta^{\Delta}(k)= \frac{i}{k^0 -\Delta +i \epsilon}
\label{propNDtext}
\end{equation}
We see that the difference between nucleon and $\Delta$ propagators lies in the factor $\Delta$ (of order $1/N_c$) in the denominator. It is exactly the difference that makes $\widetilde{G}_M^{I=0}$ and $\widetilde{G}_E^{I=1}$ nonzero and of order $1/N_c$.

Having all necessary ingredients (\ref{vertexFPPtext})-(\ref{propNDtext}) one can proceed to the calculation of photon-nucleon interaction amplitudes $M_{fi}$, and, from here, extract the appropriate form factors. Note  that the Fourier transformed (i.e. position-space) form factors are finite even in the chiral limit $m_{\pi}\rightarrow 0$. Moreover, sending $m_{\pi}\rightarrow 0$ allows us to perform both the loop momentum integration and the Fourier transform analytically and thus obtain the form factors in a closed form.

 The Feynman diagrams contributing to the isoscalar current are summarized in  Fig.\ref{isoscadiagrams} and described in the appendix.
Evaluating these diagrams, Fourier transforming, setting the pion mass to zeros and then
extracting the longest distance part of isoscalar electric and magnetic form factors yields:
\begin{eqnarray}
\lim\limits_{r\rightarrow \infty}\widetilde{G}_E^{I=0} & = &\frac{3^3}{2^{9} \pi^5}\, \frac{1}{f_{\pi}^3} \left(\frac{g_A}{f_{\pi}} \right)^3 \, \frac{1}{r^9} \label{resultISE},\\
\lim\limits_{r\rightarrow \infty}\widetilde{G}_M^{I=0} & =  & \frac{3}{2^{9} \pi^5}\, \frac{1}{f_{\pi}^3}\left(\frac{g_A}{f_{\pi}} \right)^3 \, \frac{\Delta}{r^7} .\label{resultISM}
\end{eqnarray}

\begin{figure}
\begin{center}

\includegraphics[width=2in]{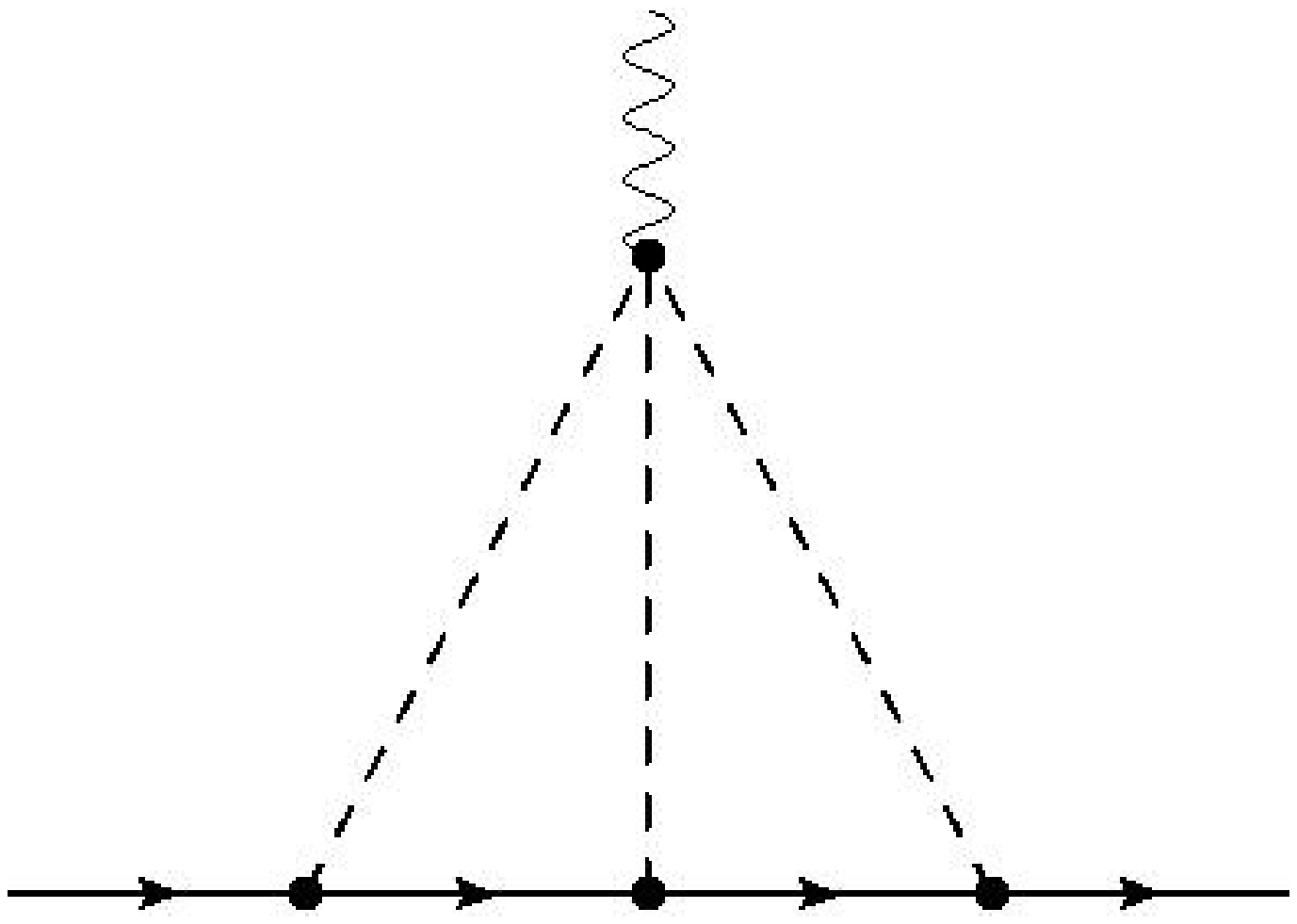}
\vskip0.1in
\includegraphics[width=2in]{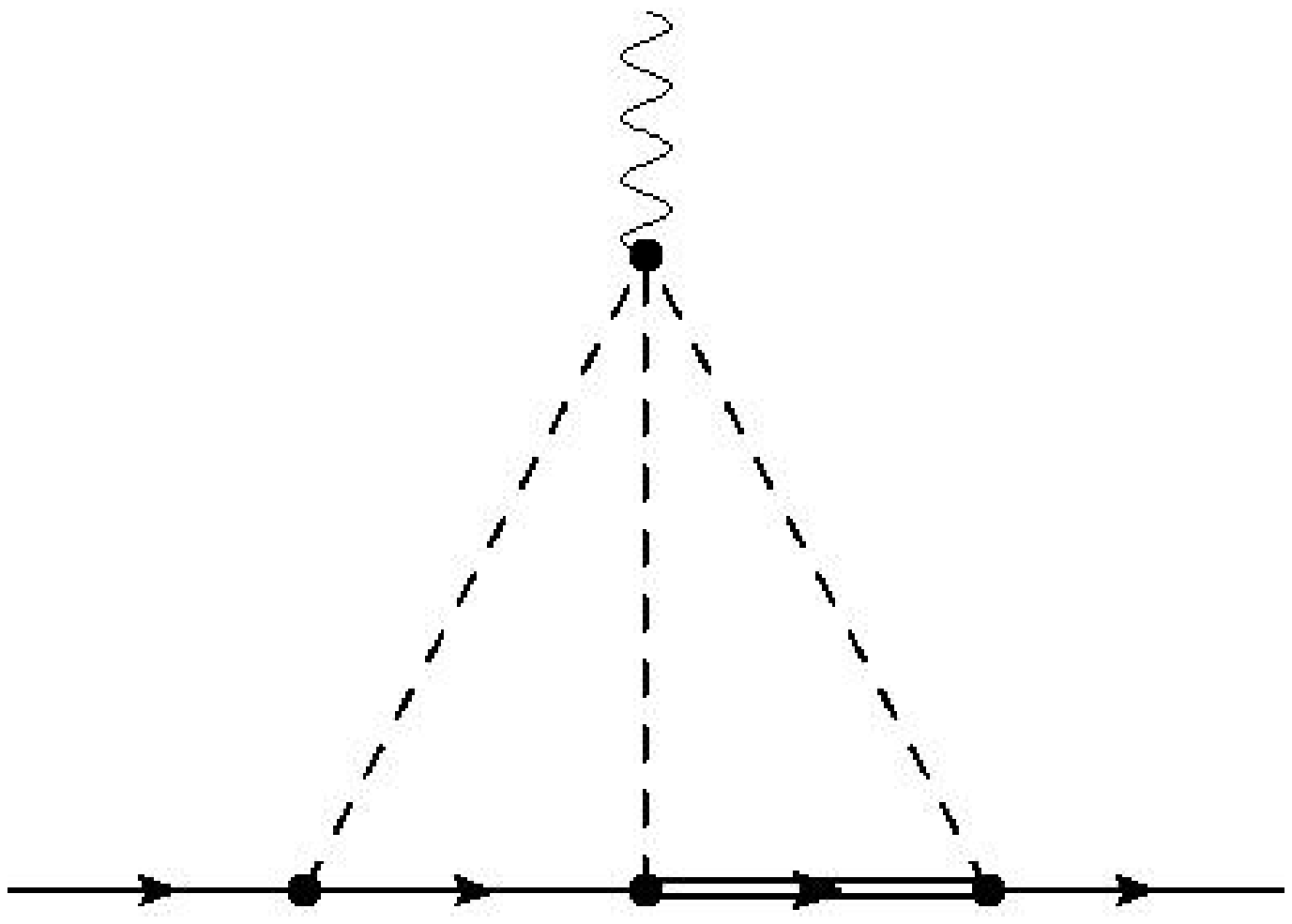}
\vskip0.1in
\includegraphics[width=2in]{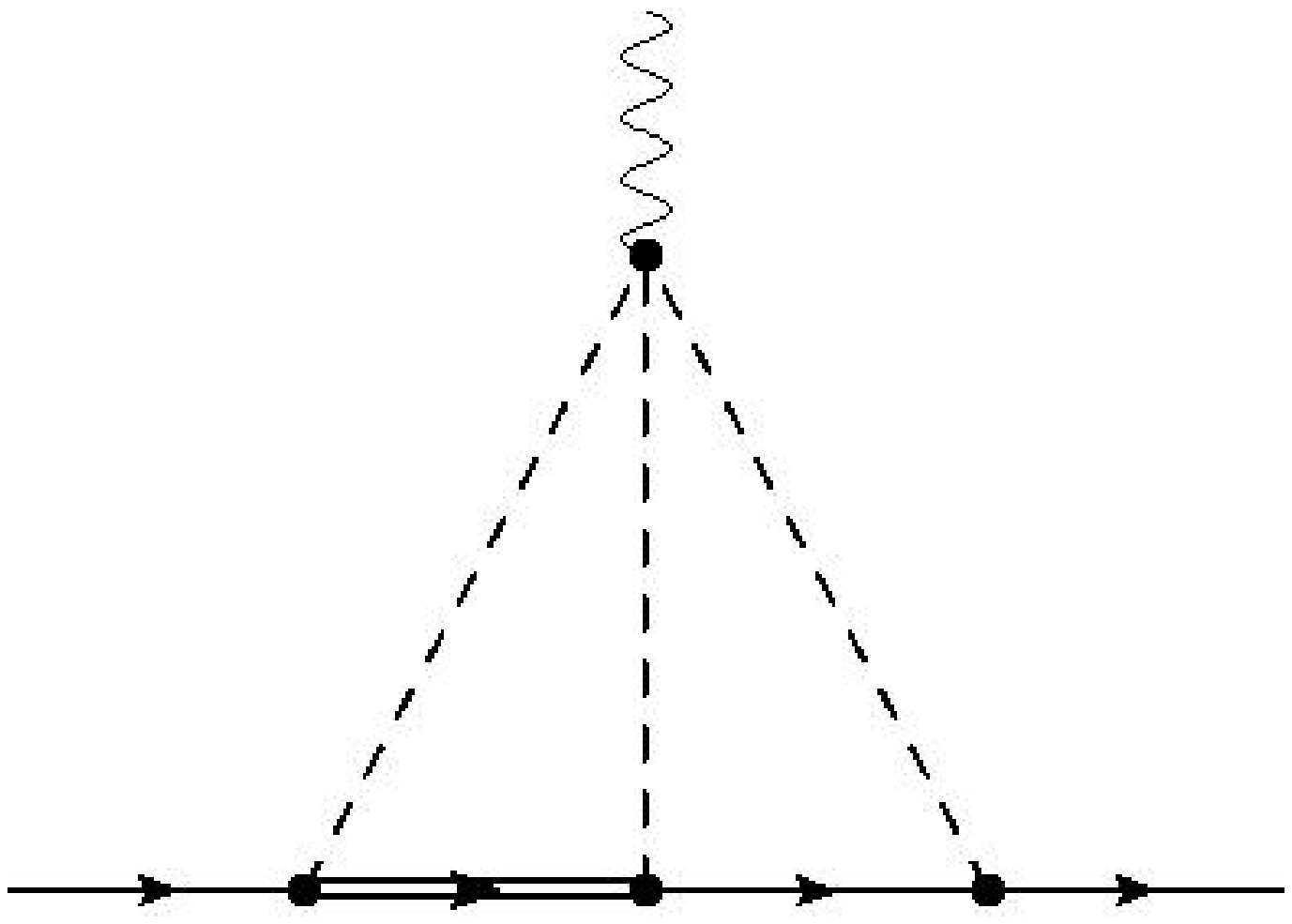}
\vskip0.1in
\includegraphics[width=2in]{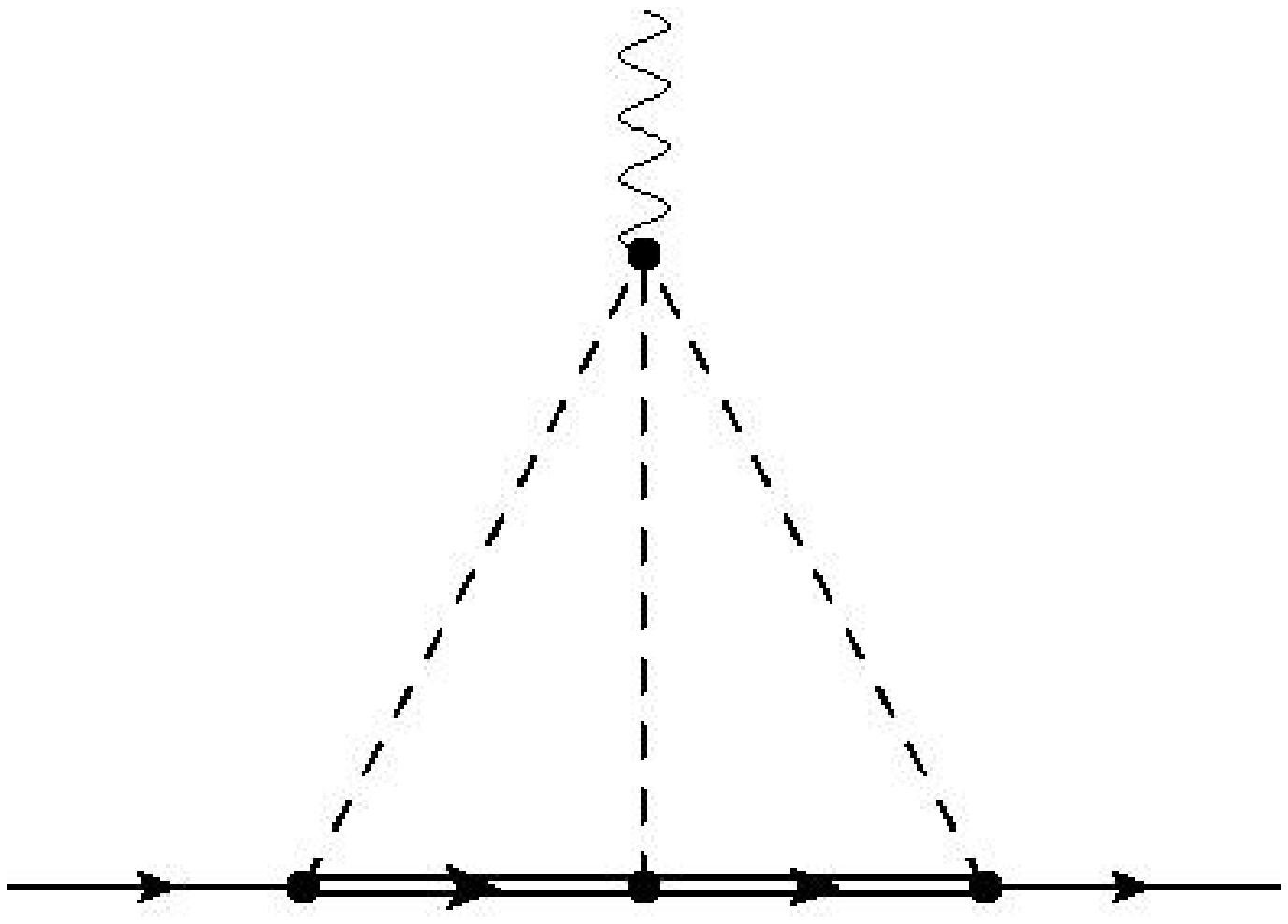}

\caption{Feynman diagrams contributing to the long distance part of the isoscalar form factors (double lines in the intermediate states represent the $\Delta$-isobar).}
\label{isoscadiagrams}

\end{center}
\end{figure}

 Feynman diagrams contributing to the isovector current are summarized in   Fig.\ref{isovecdiagrams} and also described in the appendix.
Evaluating these diagrams, Fourier transforming, setting the pion mass to zeros and then
extracting the longest distance part of isoscalar electric and magnetic form factors yields:
\begin{eqnarray}
\lim\limits_{r\rightarrow \infty} \widetilde{G}_E^{I=1} & = & \frac{1}{2^4 \pi^2}\, \left(\frac{g_A}{f_{\pi}} \right)^2 \, \frac{\Delta}{r^4} ,\label{resultIVE}\\
\lim\limits_{r\rightarrow \infty} \widetilde{G}_M^{I=1} & = & \frac{1}{2^5 \pi^2}\, \left(\frac{g_A}{f_{\pi}} \right)^2 \, \frac{1}{r^4} . \label{resultIVM}
\end{eqnarray}

\begin{figure}
\begin{center}

\includegraphics[width=2in]{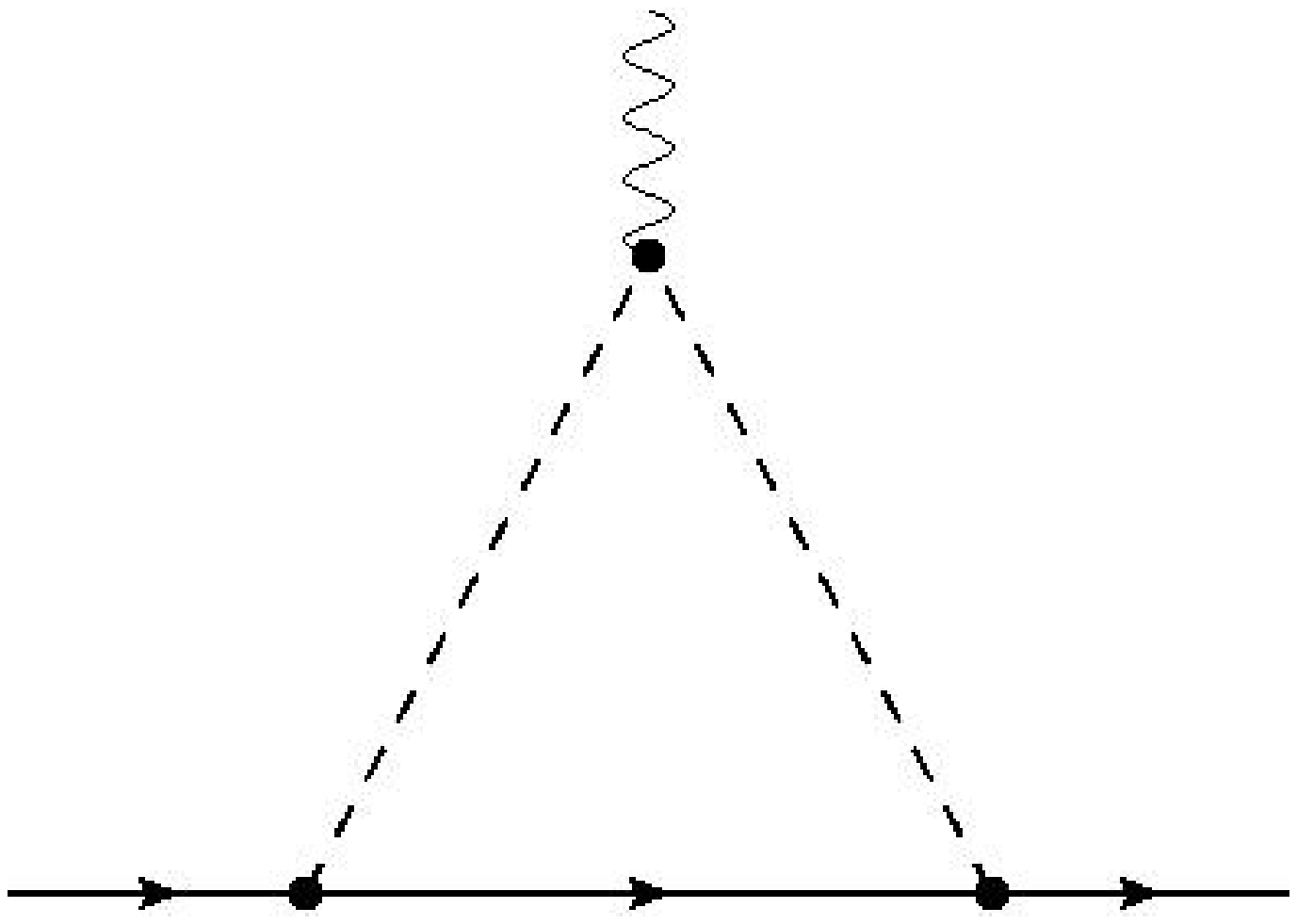}
\vskip0.1in
\includegraphics[width=2in]{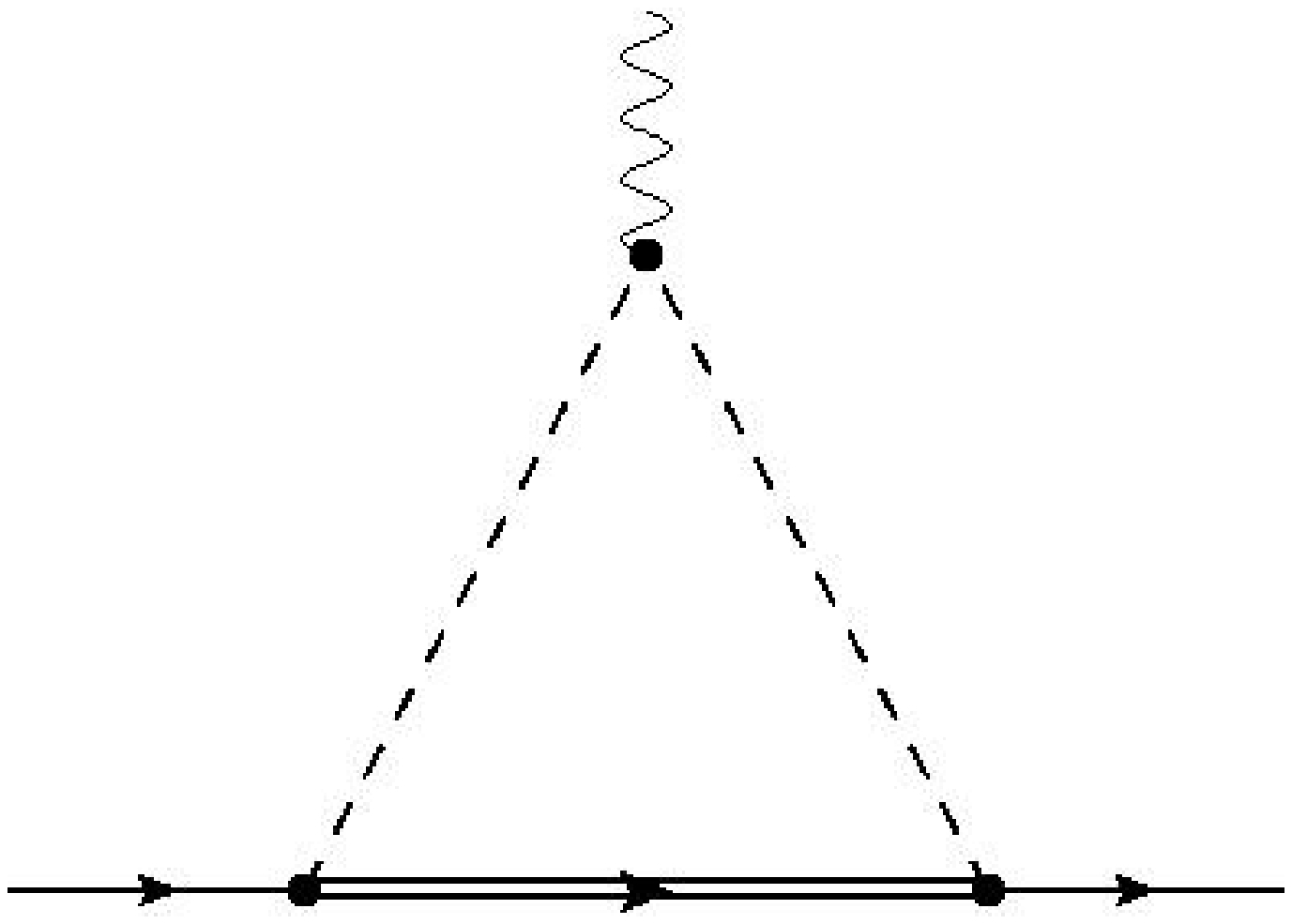}

\caption{Feynman diagrams contributing to the long distance part of the isovector form factors (double line represents the $\Delta$-isobar).}
\label{isovecdiagrams}

\end{center}
\end{figure}

Note that Eqs.~(\ref{resultISE})-(\ref{resultIVM}) are identical to those from the Skyrme model \cite{ChermanCohenNielsen}.  We have shown, however, that  these are  truly model-independent relations and so do not depend in any way on the Skyrme  model.  Of course, in verifying these results for a given model, one must use the values of $g_A$, $f_\pi$, and $\Delta$ from the same  model.  We note again that in deriving these relations we explicitly took the large $N_c$ limit prior to any others.

As noted in Ref.~\cite{ChermanCohenNielsen} one can trivially use the results of (\ref{resultISE})-(\ref{resultIVM}) to show  that the relation (\ref{magicalratio}) is satisfied in the large $N_c$ $\chi$PT.
\begin{equation}
\lim\limits_{r \rightarrow \infty} \frac{ r^2 \, \widetilde{G}_E^{I=0} \,\widetilde{G}_E^{I=1}}{\widetilde{G}_M^{I=0} \,\widetilde{G}_M^{I=1}} =18.
\label{magicalratioresult}
\end{equation}

As advertised, all low energy constants canceled in the ratio of form factors and only universal number 18 remains.
The cancellation of $\Delta$ was pointed out earlier in this letter.
However,  it is trivial to see that $g_A$ and $f_\pi$ also cancel.  This a particularly useful model-independent result.

\section{The role of the limits}

 The calculation reported above was done by taking the chiral limit---albeit after the large $N_c$ limit is taken.  It is useful to consider what happens when $m_\pi$ remains finite.  Note that the amplitude for
 finding a pion far away from the baryon is Yuakawa-like and   modifies the simple power law behavior by $\exp (-m_\pi r)$. Since the longest-range  isovector part of the electromagnetic  current is dominated by the two-pion contribution, one expects the longest range part of both the electric and magnetic  isovector  form factors to be proportional to  $\exp(-2 m_{\pi}r)$. Similarly, both the electric and magnetic isoscalar form factors are expected to be proportional to $\exp(-3 m_{\pi}r)$, since the isoscalar physics is dominated by the three-pion contribution. Thus, it appears that all terms proportional to exponentials cancel in the ratio of Eq.~(\ref{magicalratio}) and its value remains $18 $ even away from the chiral limit.

 The ratio computed in Eq.~(\ref{magicalratioresult}) in previous paragraphs was calculated by taking  $N_c \rightarrow \infty$ limit at the outset.  Note that the problem has three control parameters---$N_c$, $m_\pi$  and $r$ which is used to select out the the long distance limit $r \rightarrow \infty$  (which  is essential in our study of  the ratio of form factors).  As stressed earlier, the limits considered in our paper do not commute in general and therefore the ordering of limits must be specified.   The issue we wish to discuss here is what happens if one takes the large $N_c$ limit at the end of the problem rather than at the outset.

The key issue to address is the presence of $\Delta$-isobar in the intermediate state  since this is the origin of  the difference.  The key point is that if the large $N_c$ limit is taken at the outset, the $\Delta$ is degenerate with the nucleon, and intermediate states containing $\Delta$s contribute to the longest distance parts of the form factors.  On the other hand, if $N_c$ is finite then the $\Delta$ is   not degenerate with the nucleon and this energy difference means that diagrams with intermediate $\Delta$s contribute a shorter range than analogous diagrams with nucleons.
 Consider, for concreteness, the isovector diagrams in Fig.\ref{isovecdiagrams}. The presence of the $\Delta$ in the second diagram leads to a suppression factor of the form $e^{-\Delta r}$ in the amplitude coming from the $\Delta$ propagator. Moreover, both amplitudes contain a factor of the form $e^{-m_{\pi} r}$ coming from pion propagators. We see that taking $\Delta \rightarrow 0$ first eliminates the difference between two amplitudes. This observation lies in the core of the argument, why both diagrams must be included if $N_c \rightarrow \infty$ is taken first.  On the other hand, in the reversed order of limits ($m_{\pi}\rightarrow 0$ first, $r \rightarrow \infty$, $N_c \rightarrow \infty$ last), the contribution of the diagram with $\Delta$ remains suppressed. Thus, only the diagram with the intermediate nucleon contributes.

For isoscalar electric and isovector magnetic form factors, diagrams add  up in the leading order, thus the difference is only in the multiplicative factor:
\begin{eqnarray}
\lim\limits_{N_c \rightarrow \infty}\,\lim\limits_{r\rightarrow \infty} \, \widetilde{G}_E^{I=0} & = &  \frac{2}{9} \, \cdot \, \lim\limits_{r\rightarrow \infty} \, \lim\limits_{N_c \rightarrow \infty} \,  \widetilde{G}_E^{I=0}  , \label{resultISE_RO}\\
\lim\limits_{N_c \rightarrow \infty}\,\lim\limits_{r\rightarrow \infty} \,   \widetilde{G}_M^{I=1} & = &   \frac{2}{3} \, \cdot \, \lim\limits_{r\rightarrow \infty} \, \lim\limits_{N_c \rightarrow \infty} \, \widetilde{G}_M^{I=1}  . \label{resultIVM_RO}
\end{eqnarray}
For isoscalar magnetic and isovector electric form factors, the diagrams originally subtracted in the leading order, and form factors were proportional to $\Delta$. Thus, the form factors in reverse ordering of limits differ not only by a multiplicative factor but also by a dimensionless quantity $1/(\Delta r)$.
\begin{eqnarray}
\lim\limits_{N_c \rightarrow \infty}\,\lim\limits_{r\rightarrow \infty} \, \widetilde{G}_M^{I=0} & = &  \frac{1}{\Delta r} \, \frac{2}{3 \pi} \, \cdot \,  \lim\limits_{r\rightarrow \infty} \, \lim\limits_{N_c \rightarrow \infty} \, \widetilde{G}_M^{I=0} , \label{resultISM_RO} \\
\lim\limits_{N_c \rightarrow \infty}\,\lim\limits_{r\rightarrow \infty} \, \widetilde{G}_E^{I=1} & = &   \frac{1}{\Delta r} \, \frac{1}{\pi} \, \cdot \, \lim\limits_{r\rightarrow \infty} \, \lim\limits_{N_c \rightarrow \infty} \,  \widetilde{G}_E^{I=1}  . \label{resultIVE_RO}
\end{eqnarray}
Combining the results of Eqs.~(\ref{resultISE_RO})-(\ref{resultIVE_RO}) we see that the ratio (\ref{magicalratio}) differs by half if the order of large $N_c$ and chiral limits is reversed:
\begin{equation}
\lim\limits_{N_c \rightarrow \infty}\,\lim\limits_{r\rightarrow \infty} \, \frac{r^2 \, \widetilde{G}_E^{I=0} \, \widetilde{G}_E^{I=1}}{\widetilde{G}_M^{I=0} \, \widetilde{G}_M^{I=1}} = 9 .
\label{magicalratioresult_RO}
\end{equation}

\section{Conclusions}

 We proved  that the relation (\ref{magicalratio}) holds in the large $N_c$ $\chi$PT, provided that the $N_c \rightarrow \infty$ limit is taken at the outset of the problem. Consequently, it may serve as an honest model-independent constraint on baryon models based on large $N_c$ and chiral physics.

\emph{Acknowledgments} We want to thank A. Cherman for very useful discussions and suggestions. This work was supported by the U.S.~Department of Energy
through grant DE-FG02-93ER-40762.

\section*{Appendix}
The details of the calculations of position-space form factors are outlined in the appendix.
The electric and magnetic form factors can be  extracted from the interaction of nucleon with photon (electromagnetic field is coupled to the vector current).
We use the following convection for indexes:
Latin indexes $a,\,b,\,\dots$ indicate  isospin, Latin indexes $i,\,j,\,\dots$ represent  components of 3-vectors, Greek indexes $\mu,\,\nu,\,\dots$ label components of Lorentz 4-vectors.

The general form of baryon current reads:
\begin{equation}
\langle \vec{p}'| J_{\mu}(q^2) | \vec{p} \rangle = \bar{u}(\vec{p}') \left( \gamma_{\mu}F_1(q^2) + \frac{i}{2M} \sigma_{\mu\nu}q^{\nu} F_2(q^2) \right) u(\vec{p}) , \nonumber
\end{equation}
where $\vec{p}'=\vec{p}+\vec{q}$ and the particles are on-shell. $F_1$ and $F_2$  are Dirac and Pauli form factors.
This expression simplifies substantially in the non-relativistic limit, which is justified by the large $N_c$ limit. The time ($0^{\rm th}$) component of the vector current can be expressed solely in term of an electric form factor $G_E$, as well as spatial  ($i^{\rm th}$) components can be expressed solely in terms of a magnetic form factor.
\begin{eqnarray}
\langle \vec{p}'| J_{0}(q^2) | \vec{p} \rangle &=& U^{\dagger} \,\, G_E(q^2) \,\, U, \nonumber\\
\langle \vec{p}'| \vec{J}_{i}(q^2) | \vec{p} \rangle &=& U^{\dagger} \,\, \frac{-i}{2M} \epsilon_{ijk} \,\vec{q}_j \, \sigma_k \, G_M(q^2) \,\, U  \label{currentsnonrel};
\end{eqnarray}
the electric and magnetic form factors $G$ are related to the original Dirac and Pauli form factors $F$ via
\begin{eqnarray}
G_E(q^2) &=& F_1(q^2) + F_2(q^2) \frac{q^2}{4M^2}, \nonumber \\
G_M(q^2) &=& F_1(q^2) + F_2(q^2) . \nonumber
\end{eqnarray}

In order to extract the position-space form factors from the momentum-space amplitudes $M_{fi}=eA_{\mu}J^{\mu}$  the convection used in Ref. \cite{ChermanCohenNielsen} was adopted:
\begin{eqnarray}
\widetilde{G}_E(r) & = &  \int \frac{{\rm d}\Omega_x}{4\pi}  \int \frac{{\rm d}^3 q}{(2 \pi)^3} \, e^{i \vec{q} \cdot \vec{x}}  \,\langle \vec{p}'| \, J^0 \,| \vec{p} \rangle , \nonumber\\
\widetilde{G}_M(r) & = & \int \frac{{\rm d}\Omega_x}{4\pi} \int \frac{{\rm d}^3 q}{(2 \pi)^3} \, e^{i \vec{q} \cdot \vec{x}} \,\frac{1}{2} \epsilon_{ij3} \,  \vec{x}_j \,\langle \vec{p}'| \,\vec{J}_i \,| \vec{p} \rangle .  \label{definitionsofFF}
\end{eqnarray}
The expansion of current matrix elements (\ref{currentsnonrel}) allows us to distinguish and define the isoscalar and the isovector current. Recall  that the wave function $U$ is not only the two-component spinor in the spin space, but also a two-component spinor in the isospin space. The most general diagonal matrix in the isospin space (incoming and outgoing particles are the same) can be written as a combination of the identity matrix ${\bf I}_{\tau}$ and the third Pauli matrix $\tau_3$. We define the term proportional to the identity matrix $U^{\dagger} \, {\bf I}_{\tau} \, U$ to be the isoscalar current (proton plus neutron) and the  term proportional to the Pauli matrix $U^{\dagger} \, \tau_3 \, U$ to be the isovector current (proton minus neutron). This definition coincides with the definitions of \cite{AdkinsNappiWitten, ChermanCohenNielsen}.

\subsection{Feynman rules}

The Feynman rule for the interaction of photon with two pions reads \cite{BernardKaiserMeissner}:
\begin{equation}
 \epsilon_{a3b} \, A_{\mu} \, (p_a^{\mu} + p_b^{\mu}),
\label{vertexFPP}
\end{equation}
where $p_a$ and $p_b$ are the incoming and outgoing pion 4-momenta, respectively.

The coupling of photon with three pions is derived from the anomalous baryon current in QCD
\cite{Witten82}.
The Feynman rule for the corresponding vertex reads:
\begin{equation}
\frac{1}{12 \pi^2 f_{\pi}^3} \, \epsilon_{abc} \, \epsilon^{\mu \nu \kappa \lambda }   A_{\mu} \, p_{a{\nu}} \, p_{b{\kappa}} \, p_{c{\lambda}},
\label{vertexFPPP}
\end{equation}
where all momenta are outgoing.

The interaction of pion with baryons is of the vector-isovector form and is derivatively coupled. The Feynman rule for incoming baryon $B$ and outgoing baryon $B'$ reads
\begin{equation}
\frac{g_A}{2 f_{\pi}  } \,\,\sqrt{\frac{2J^{(B')}+1}{2J^{(B)}+1}} \,\, \tau^{(BB')}_a \,\, \sigma^{(BB')}_i \, \vec{p}_i,
\label{vertexPBB}
\end{equation}
where $\vec{p}_i$ is the outgoing 3-momentum of a pion with isospin $a$.
In (\ref{vertexPBB}), $\tau^{BB'}$ and $\sigma^{BB'}$ are operators acting in the isospin and in the spin space, respectively.  The form of the coupling  is derived from the consistency relations of large $N_c$ QCD \cite{DashenManohar93-1, DashenManohar93-2}; they require not only the degeneracy of the whole tower of states with $I=J$ (nucleon with $I=J=1/2$, $\Delta$ with $I=J=3/2$, \dots), but also specify the pion-baryon-baryon$'$ vertex. For example, the matrix elements for incoming nucleon and outgoing delta reads:
\begin{equation}
\left( \tau^{(N\Delta)}_a \right)_{\alpha\alpha'} = \sqrt{3} \, \left(\frac{1}{2}\,\alpha\,\,,\,\,1\,a\,|\,\frac{3}{2}\,\alpha' \right). \nonumber
\end{equation}
All the matrices we need read:
\begin{eqnarray}
\tau_1,\sigma^{(NN)}_1 &=& \tau_1,\sigma_1 =\left( \begin{array}{cc} 0 & 1 \\ 1 & 0 \end{array} \right) ,\nonumber \\
\tau_2,\sigma^{(NN)}_2 &=& \tau_2,\sigma_2 =i \left( \begin{array}{cc} 0 & -1 \\ 1 & 0 \end{array} \right) ,\nonumber \\
\tau_3,\sigma^{(NN)}_3 &=& \tau_3,\sigma_3 =\left( \begin{array}{cc} 1 & 0 \\ 0 & -1 \end{array} \right), \nonumber
\end{eqnarray}

\begin{eqnarray}
\tau_1,\sigma^{(N\Delta)}_1 &=& \left( \begin{array}{cc} -\sqrt{\frac{3}{2}} & 0 \\ 0 & -\sqrt{\frac{1}{2}} \\ \sqrt{\frac{1}{2}} & 0 \\ 0 & \sqrt{\frac{3}{2}} \end{array} \right), \nonumber \\
\tau_2,\sigma^{(N\Delta)}_2 &=& i \left( \begin{array}{cc}   \sqrt{\frac{3}{2}} & 0 \\ 0 &  \sqrt{\frac{1}{2}} \\  \sqrt{\frac{1}{2}} & 0 \\ 0 & \sqrt{\frac{3}{2}} \end{array} \right) ,\nonumber \\
\tau_3,\sigma^{(N\Delta)}_3 &=& \left( \begin{array}{cc} 0 & 0 \\  \sqrt{2} &0 \\ 0& \sqrt{2}  \\ 0 & 0 \end{array} \right) ,\nonumber
\end{eqnarray}

\begin{eqnarray}
\tau_1,\sigma^{(\Delta N)}_1 &=& \left( \begin{array}{cccc} \frac{\sqrt{3}}{2} & 0 & -\frac{1}{2} & 0 \\ 0 & \frac{1}{2} & 0 & -\frac{\sqrt{3}}{2} \end{array} \right) ,\nonumber \\
\tau_2,\sigma^{(\Delta N)}_2 &=& i \left( \begin{array}{cccc}   \frac{\sqrt{3}}{2} & 0 &   \frac{1}{2} & 0 \\ 0 &   \frac{1}{2} & 0 &   \frac{\sqrt{3}}{2} \end{array} \right) ,\nonumber \\
\tau_3,\sigma^{(\Delta N)}_3 &=& \left( \begin{array}{cccc} 0 & -1 & 0 &0  \\ 0 & 0& -1 &0  \end{array} \right) ,\nonumber
\end{eqnarray}

\begin{eqnarray}
\tau_1,\sigma^{(\Delta \Delta)}_1 &=& \left( \begin{array}{cccc} 0& \sqrt{\frac{3}{5}} & 0 & 0 \\    \sqrt{\frac{3}{5}} & 0 &  \frac{2}{\sqrt{5}} &0 \\ 0& \frac{2}{\sqrt{5}}&0& \sqrt{\frac{3}{5}} \\ 0&0& \sqrt{\frac{3}{5}}&0 \end{array} \right) ,\nonumber \\
\tau_2,\sigma^{(\Delta \Delta)}_2 &=& i \left( \begin{array}{cccc} 0& - \sqrt{\frac{3}{5}} & 0 & 0 \\    \sqrt{\frac{3}{5}} & 0 & -  \frac{2}{\sqrt{5}} &0 \\ 0&   \frac{2}{\sqrt{5}}&0& -  \sqrt{\frac{3}{5}} \\ 0&0&   \sqrt{\frac{3}{5}}&0 \end{array} \right) ,\nonumber \\
\tau_3,\sigma^{(\Delta \Delta)}_3 &=& \left( \begin{array}{cccc}  \frac{3}{\sqrt{5}} & 0 & 0 & 0 \\  0& \frac{1}{\sqrt{5}} & 0 &0 \\ 0&0&-\frac{1}{\sqrt{5}}& 0 \\ 0&0& 0&-\frac{3}{\sqrt{5}} \end{array} \right) .\nonumber
\end{eqnarray}
Note that the matrices for the nucleon-nucleon vertex are ordinary Pauli matrices. Thus, the general form (\ref{vertexPBB}) of the pion-baryon-baryon$'$ reproduces the standard $\chi$PT \cite{BernardKaiserMeissner} in the pion-nucleon-nucleon case.

The other three building blocks are the propagators of particles in the intermediate states.
Since we intend to set $m_{\pi} \rightarrow 0$ in the end (chiral limit) we must use the full relativistic pion propagator ($k$ being propagating 4-momentum):
\begin{equation}
\Delta^{\pi}(k)= \frac{i}{k^2-m_{\pi}^2+ i \epsilon} .
\label{propP}
\end{equation}
The heavy baryon approximation, which is justified by the fact that we work in the large $N_c$ limit, allows us to treat baryons non-relativistically:
\begin{equation}
\Delta^{N}(k)= \frac{i}{k^0+i \epsilon}\,,\,\,\,\Delta^{\Delta}(k)= \frac{i}{k^0 -\Delta +i \epsilon} .
\label{propND}
\end{equation}
We see that the difference between nucleon and $\Delta$ propagators lies in the factor $\Delta$ (of order $1/N_c$) in the denominator. It is exactly the difference that makes $\widetilde{G}_M^{I=0}$ and $\widetilde{G}_E^{I=1}$ nonzero and of order $1/N_c$.

\subsection{Isovector form factors}
First, we present the calculation of isovector form factors. They result from Feynman diagrams containing one loop. Therefore, their calculation is  simpler and easier to follow than the calculation of isoscalar ones.

From the building blocks (\ref{vertexFPP}), (\ref{vertexPBB}), (\ref{propP}), (\ref{propND}), one can straightforwardly construct the amplitudes corresponding to the diagrams in Fig.  \ref{isovecdiagrams}.
They can be split into two parts: one that is the same for both diagrams, and one that is different.
The difference lies in the matrices involved in pion-baryon-baryon$'$ vertex and in the propagators of intermediate baryons.

The total isovector amplitude including the combinatoric factor $2!$ is:
\begin{widetext}
\begin{eqnarray}
M_{fi} = e A_{\mu}J^{\mu}_{I=1} &=&   e A_{\mu} \,\, 4 i  \, \left( \frac{g_A}{2 f_{\pi}} \right)^2 \,  \epsilon_{ab3}    \int \frac{{\rm d}^4 k}{(2\pi)^4}   k^{\mu} \left(\frac{\vec{q}}{2} +\vec{k} \right)_l \left(\frac{\vec{q}}{2} -\vec{k} \right)_n  \  \Delta^{\pi} \left(k+\frac{q}{2} \right) \,\, \Delta^{\pi} \left(k-\frac{q}{2} \right)    \nonumber \\
 & &  U^{\dagger} \left( \tau^{(NN)}_b \tau^{(NN)}_a   \,\, \sigma^{(NN)}_n \sigma^{(NN)}_l \,\,  \Delta^{N}(k)  \,\, +\,\,  \tau^{(\Delta N)}_b \tau^{(N\Delta)}_a  \,\, \sigma^{(\Delta N)}_n \sigma^{(N\Delta)}_l \,\,   \Delta^{\Delta}(k)   \right) U.  \label{amplitudeisovecdiffer}
\end{eqnarray}
\end{widetext}

The product of two matrices (\ref{amplitudeisovecdiffer}) can be decomposed into two parts: first being proportional to identity matrix, and second proportional to the Pauli matrix. The products of two isospin matrices read:
\begin{eqnarray}
\tau^{(NN)}_a \tau^{(NN)}_b &=& \delta_{ab} \, {\bf I}_{\tau} + i\,  \epsilon_{abc} \, \tau_c ,\nonumber \\
\tau^{(\Delta N)}_a \tau^{(N\Delta)}_b &=&  -\sqrt{2} \, \delta_{ab} \, {\bf I}_{\tau} + \frac{i}{\sqrt{2}} \epsilon_{abc} \, \tau_c , \label{productoftwo}
\end{eqnarray}
with   analogous relations for spin matrices $\sigma$.

In the isospin space, only the part proportional to $\tau_3$ contributes due to the factor $\epsilon_{ab3}$, which originates in the photon-2 pion vertex. It leads to an  overall factor $i \epsilon_{ab3} \epsilon_{bac} \tau_c = -2 i \tau_3$.
Recall that the matrix element where the isospin part of the nucleon wave functions are coupled via $\tau_3$ matrix represents  the isovector current.

In the spin space, we must take into account both scalar (proportional to ${\bf I}_{\sigma}$) and vector (proportional to $\sigma_i$) matrix elements.
The terms proportional to ${\bf I}_{\sigma}$ cancel completely if nucleon and $\Delta$ propagators are equal, $\Delta \rightarrow 0$. Thus, one must include the $N$-$\Delta$ mass splitting and the isovector-scalar baryon matrix element is proportional to $\Delta \sim 1/N_c$.
On the other hand, the terms proportional to $\sigma_i$ add together and the $N$-$\Delta$ mass splitting can be ignored.
Then,  (\ref{amplitudeisovecdiffer}) can be rewritten as

\begin{equation}
U^{\dagger} \, 2 \, \tau_3  \left(  {\bf I}_{\sigma}  \, \delta_{nl} \, \frac{-\Delta}{(k^0 + i \epsilon)^2 } +    \sigma_3   \,  \epsilon_{nl3} \, \frac{3i}{2(k^0 +i \epsilon) }  \right) U .\nonumber
\end{equation}

The definition  of the position-space form factors (\ref{definitionsofFF}) guarantees that only the scalar part (proportional to ${\bf I}_{\sigma}$) contributes to the electric form factor $\widetilde{G}_E$ and only the vector part  (proportional to $\sigma_i$) contributes to the magnetic form factor $\widetilde{G}_M$. This is because only the scalar part survives the overall angular integration.

The electric form factor is the Fourier transform of the zeroth component of the current:
\begin{eqnarray}
\widetilde{G}_E^{I=1} &=&  8 i  \Delta  \left( \frac{g_A}{2 f_{\pi}} \right)^2 U^{\dagger}\tau_3 {\bf I}_{\sigma} U \, \int \frac{{\rm d}^3 q}{(2 \pi)^3} \, e^{i \vec{q} \cdot \vec{x}} \nonumber \\
&&  \int \frac{{\rm d}^4 k}{(2 \pi)^4} \, \left( \frac{\vec{q}}{2}  +\vec{k} \right)\cdot\left(\frac{\vec{q}}{2} -\vec{k} \right)  \,\frac{k^0}{(k^0+i\epsilon)^2} \nonumber \\
&& \Delta^{\pi}\left(k+\frac{q}{2}\right) \,\, \Delta^{\pi}\left(k-\frac{q}{2}\right) \nonumber \\
& \underset{r\rightarrow \infty}{\underset{m_{\pi}\rightarrow 0}{=}} & \frac{1}{2^4 \pi^2}\, \left(\frac{g_A}{f_{\pi}} \right)^2 \, \frac{\Delta}{r^4} ,
\end{eqnarray}
with the normalization $U^{\dagger}U=1$ used.

Analogously, the magnetic form factor comes from the Fourier transform of the spatial component of currents:
\begin{eqnarray}
\widetilde{G}_M^{I=1} &=&  6  \left( \frac{g_A}{2 f_{\pi}} \right)^2 U^{\dagger}\tau_3 \sigma_3 U     \int \frac{{\rm d}\Omega_x}{4\pi} \int \frac{{\rm d}^3 q}{(2 \pi)^3}  e^{i \vec{q} \cdot \vec{x}}   \nonumber \\
&& \int \frac{{\rm d}^4 k}{(2 \pi)^4} \, \epsilon_{ij3} \vec{x}_j \vec{k}_i \, \epsilon_{nl3} \left(\frac{\vec{q}}{2} +\vec{k} \right)_l \left(\frac{\vec{q}}{2} -\vec{k}\right)_n \nonumber \\
&& \frac{1}{k^0+ i\epsilon} \,\, \Delta^{\pi}\left(k+\frac{q}{2}\right) \,\, \Delta^{\pi}\left(k-\frac{q}{2}\right) \nonumber \\
& \underset{r\rightarrow \infty}{\underset{m_{\pi}\rightarrow 0}{=}} & \frac{1}{2^5 \pi^2}\, \left(\frac{g_A}{f_{\pi}} \right)^2 \, \frac{1}{r^4} .
\end{eqnarray}

\vskip0.5in
\subsection{Isoscalar form factors}
The isoscalar form factors are derived from Feynman diagrams including two loops, see Fig. \ref{isoscadiagrams}. Even though the presence of two loop momenta, $k$ and $l$, makes the integrals more complicated from the mathematical point of view,  the basic ideas remain  the same as earlier.  The total isoscalar amplitude including the combinatoric factor 3! reads:

\begin{widetext}
\begin{eqnarray}
M_{fi} = e A_{\mu}J^{\mu}_{I=0} &=& e A_{\mu}  \frac{i}{2 \pi^2 f_{\pi}^3}  \, \left( \frac{g_A}{2 f_{\pi}} \right)^3 \, \epsilon_{abc}    \int \frac{{\rm d}^4 k}{(2\pi)^4} \int \frac{{\rm d}^4 l}{(2\pi)^l} \,\,\Delta^{\pi}\left(k+\frac{q}{2}\right) \,\, \Delta^{\pi}\left(k+l\right) \,\, \Delta^{\pi}\left(l+\frac{q}{2}\right)   \nonumber\\
&&  \epsilon^{\mu \alpha \beta \gamma}    \left(k+\frac{q}{2}\right)_{\alpha} \left(-k-l\right)_{\beta} \left(l+\frac{q}{2}\right)_{\gamma}   \left(\vec{k}+\frac{\vec{q}}{2}\right)_r \left(-\vec{k}-\vec{l}\right)_s \left(\vec{l}+\frac{\vec{q}}{2}\right)_t   \nonumber \\
&  &  U^{\dagger} \left(  \tau^{(NN)}_c \tau^{(NN)}_b \tau^{(NN)}_a        \sigma^{(NN)}_t \sigma^{(NN)}_s \sigma^{(NN)}_r   \Delta^{N}(k) \Delta^{N}(-l) \,\,\,  + \right.  \nonumber \\
&& \,\,\,  \left. \tau^{(\Delta N)}_c \tau^{(N\Delta)}_b \tau^{(NN)}_a       \sigma^{(\Delta N)}_t \sigma^{(N\Delta)}_s \sigma^{(NN)}_r   \Delta^{N}(k) \Delta^{\Delta}(-l)  + \dots  \right) \,\, U,
\end{eqnarray}

where the ellipsis stands for terms corresponding to the remaining diagram with one $\Delta$ in the intermediate state and to the diagram with two $\Delta$s in the intermediate state.
The  product of three matrices simplifies in a spirit similar to Eqs. (\ref{productoftwo}) to a piece proportional to identity matrix and to a piece proportional to Pauli matrix  (only isospin matrices $\tau$ are shown, since both spin $\sigma$ are isospin $\tau$ matrices are the same):

\begin{eqnarray}
\tau^{(NN)}_a \tau^{(NN)}_b \tau^{(NN)}_c &=& i \epsilon_{abc} \, {\bf I}_{\tau} +   \delta_{ab} \tau_{c} -   \delta_{ac} \tau_{b} +  \delta_{bc} \tau_{a} , \nonumber\\
\tau^{(\Delta N)}_a \tau^{(N\Delta)}_b \tau^{(NN)}_c &=& \frac{i}{\sqrt{2}} \epsilon_{abc} \, {\bf I}_{\tau} -\sqrt{2}  \delta_{ab} \tau_{c}  - \frac{1}{\sqrt{2}}  \delta_{ac} \tau_{b} + \frac{1}{\sqrt{2}} \delta_{bc} \tau_{a} ,\nonumber\\
\tau^{(NN)}_a \tau^{(\Delta N)}_b \tau^{(N\Delta)}_c &=& \frac{i}{\sqrt{2}} \epsilon_{abc} \, {\bf I}_{\tau} + \frac{1}{\sqrt{2}} \delta_{ab} \tau_{c}   - \frac{1}{\sqrt{2}}  \delta_{ac} \tau_{b} - \sqrt{2}  \delta_{bc} \tau_{a} , \nonumber\\
\tau^{(\Delta N)}_a \tau^{(\Delta \Delta)}_b \tau^{(N\Delta)}_c &=& -i \sqrt{\frac{5}{2}} \epsilon_{abc} \, {\bf I}_{\tau} + \frac{1}{\sqrt{10}} \delta_{ab} \tau_{c} -  2\sqrt{\frac{2}{5}} \delta_{ac} \tau_{b} + \frac{1}{\sqrt{10}} \delta_{bc} \tau_{a} .\label{prouctofthree}
\end{eqnarray}
\end{widetext}

Unlike to the isovector amplitude,  only the part proportional to the identity matrix in the isospin space contributes (recall $\epsilon_{abc}$ from photon-3 pions vertex) to the isoscalar amplitude.
The overall factor $i \epsilon_{abc} \epsilon_{cba} {\bf I}_{\tau} = -6 i {\bf I}_{\tau}$ emerges.

In the spin space, the scalar parts (contributing to $G_E$) add together and yield an overall multiplicative factor $\frac{9}{2} {\bf I}_{\sigma}$. On the other hand, vector parts  cancel exactly if $N$-$\Delta$ mass splitting is neglected. Thus, one must include $\Delta$ in the calculations. Recall that the same thing happened to the electric isovector form factor. It is in agreement with the work of Broniowski  and  Cohen \cite{CohenBroniowski, Cohen96}, who showed that vector-isoscalar and scalar-isovector nucleon matrix elements are zero in the leading order of $1/N_c$ expansion.

The electric form factor is the Fourier transform of the zeroth component of the current:
\begin{eqnarray}
\widetilde{G}_E^{I=0} &=& \frac{- 27}{2 \pi^2 f_{\pi}^3}  \left( \frac{g_A}{2 f_{\pi}} \right)^3 U^{\dagger}{\bf I}_{\tau} {\bf I}_{\sigma} U \, \int \frac{{\rm d}^3 q}{(2 \pi)^3} \frac{{\rm d}^4 k}{(2 \pi)^4}\frac{{\rm d}^4 l}{(2 \pi)^4}\nonumber \\
&&  e^{i \vec{q} \cdot \vec{x}} \,\left( \epsilon_{rst} \, \vec{k}_r \vec{l}_s \vec{q}_t \right)^2 \, \Delta^{N}(k) \Delta^{N}(-l) \nonumber \\
&& \Delta^{\pi}\left(k+\frac{q}{2}\right) \,\, \Delta^{\pi}\left(k+l\right) \,\, \Delta^{\pi}\left(l+\frac{q}{2}\right) \nonumber \\
& \underset{r\rightarrow \infty}{\underset{m_{\pi}\rightarrow 0}{=}} &  \frac{3^3}{2^{9} \pi^5}\, \frac{1}{f_{\pi}^3} \left(\frac{g_A}{f_{\pi}} \right)^3 \, \frac{1}{r^9} .
\end{eqnarray}

The calculation of isoscalar magnetic form factor is of the same spirit. However, it is much longer and more tedious than previous ones, so only a brief sketch of the procedure is shown.
Note, that the magnetic form factor is derived from the spatial components of the baryon current $J^{\mu=i}$. There are three similar terms coming from the four-dimensional Levi-Civita tensor $\epsilon^{i \alpha \beta \gamma}$, where either $\alpha$ or $\beta$ or $\gamma$ can be of value 0. Moreover, from the decomposition of products of three matrices (\ref{prouctofthree}), there emerge three options how the products of momenta are arranged (how the indices of variables $k$, $l$, and $q$ are  matched). Thus there are nine individual contributions to the isoscalar magnetic form factor. Combining these yields:
\begin{equation}
\widetilde{G}_M  \underset{r\rightarrow \infty}{\underset{m_{\pi}\rightarrow 0}{=}}    \frac{3}{2^{9} \pi^5}\, \frac{1}{f_{\pi}^3} \left(\frac{g_A}{f_{\pi}} \right)^3 \, \frac{\Delta}{r^7} .
\end{equation}

\end{document}